\documentclass[conference]{IEEEtran}
\usepackage{graphicx}
\usepackage{eurosym}
\usepackage{amssymb}
\usepackage{amsmath}
\usepackage{amsfonts}
\usepackage{epstopdf}
\usepackage{epsf,subfigure}
\usepackage{psfrag}
\usepackage{graphics}
\usepackage{color} 
\usepackage{hyperref}
\usepackage{cite}
\usepackage{array,multirow,pbox}
\usepackage{caption}
\usepackage[verbose]{placeins}

\newcommand{\rem}[1]{}

\newtheorem{definition}{Definition}

\newtheorem{remark}{Remark}

\newcommand{\qed}{\nobreak \ifvmode \relax \else
      \ifdim\lastskip<1.5em \hskip-\lastskip
      \hskip1.5em plus0em minus0.5em \fi \nobreak
      \vrule height0.75em width0.5em depth0.25em\fi}

\graphicspath{{Figures/}}

\IEEEoverridecommandlockouts

\ifCLASSINFOpdf
\else
\fi
\begin{document}
%
\title{A Nonlinear Regression Method for Composite Protection Modeling of Induction Motor Loads}

\author{\IEEEauthorblockN{Soumya Kundu\IEEEauthorrefmark{1},  Zhigang Chu\IEEEauthorrefmark{2}, Yuan Liu\IEEEauthorrefmark{1}, Yingying Tang\IEEEauthorrefmark{3}, Qiuhua Huang\IEEEauthorrefmark{1}, Daniel James\IEEEauthorrefmark{1}, Yu Zhang\IEEEauthorrefmark{4}, \\
Pavel Etingov\IEEEauthorrefmark{1} and David. P. Chassin\IEEEauthorrefmark{5}}
\IEEEauthorblockA{\IEEEauthorrefmark{1}Pacific Northwest National Laboratory, USA\\Email:\,\{soumya.kundu,\,yuan.liu,\,qiuhua.huang,\,daniel.james,
pavel.etingov\}@pnnl.gov,}
\IEEEauthorblockA{\IEEEauthorrefmark{2}Arizona State University, USA. Email: zhu2@asu.edu,}
\IEEEauthorblockA{\IEEEauthorrefmark{3}Microsoft Corporation, USA. Email: yita@microsoft.com,}
\IEEEauthorblockA{\IEEEauthorrefmark{4}Pacific Gas and Electric Company, USA. Email: yu.zhang@pge.com,}
\IEEEauthorblockA{\IEEEauthorrefmark{5}SLAC National Accelerator Laboratory, USA. Email: dchassin@slac.stanford.edu}%
\thanks{Authors gratefully acknowledge the support provided by U.S. Department of Energy Office of Electricity Delivery \& Energy Reliability for this work which was carried out at Pacific Northwest National Laboratory (under contract DE-AC02 -76RL01830) and SLAC National Accelerator Laboratory (under contract DE-AC06-76SF00515). Authors Zhigang Chu, Yingying Tang and Yu Zhang were with Pacific Northwest National Laboratory when the work was performed.}}

%


\maketitle

\begin{abstract}
Protection equipment is used to prevent damage to induction motor loads by isolating them from power systems in the event of severe faults. Modeling the response of induction motor loads and their protection is vital for power system planning and operation, especially in understanding system's dynamic performance and stability after a fault occurs. Induction motors are usually equipped with several  types of protection with different operation mechanisms, making it challenging to develop adequate yet not overly complex protection models and determine their parameters for aggregate induction motor models. This paper proposes an optimization-based nonlinear regression framework to determine  protection model parameters for aggregate induction  motor loads in commercial buildings. Using a mathematical abstraction, the task of determining a suitable set of parameters for the protection model in composite load models is formulated as a nonlinear regression problem. Numerical examples are provided to illustrate the application of the framework. Sensitivity studies are presented to demonstrate the impact of lack of available motor load information on the accuracy of the protection models.
\end{abstract}


%
\IEEEpeerreviewmaketitle

\section{Introduction}
Traditionally distribution systems loads have been modeled as lumped constant impedance (Z), constant current (I), or constant power (P) loads (abbreviated as ZIP) in transmission systems studies. Some early efforts towards detailed load modeling resulted in the component-based load models using load class and composition data \cite{EPRI}, and the "interim" load model \cite{interim}. However, such load models were found to be inadequate to represent the Fault-Induced Delayed Voltage Recovery (FIDVR) phenomenon \cite {FIDVR,kevin} which are of increasing concern for the safe and secure operation of power system networks \cite{workshop}. The fact that FIDVR events are not well represented in power system studies has fueled several efforts in recent years towards the development of a "composite load model" \cite{composite,Chassin}, focusing in particular on loads with a high penetration of induction motors. A composite load model for dynamic simulations has been developed \cite {composite}, and used in both planning and operation in Western Electricity Coordinating Council (WECC) in United States. There are four types of electric motors in the WECC composite load model - 1) {motor A}: three phase ($3\phi$) induction motors that operate under constant torque. Examples of such motors include air-conditioners and refrigerators in large commercial buildings; 2) {motor B}: $3\phi$ induction motors with high inertia, operating under speed-dependent torque. Examples include fan motors in residential and commercial buildings; 3) {motor C}: $3\phi$ induction motors with low inertia, operating under speed-dependent torque. Examples include pump motors in commercial buildings; and 4) {motor D}: $1\phi$ induction motors. Examples include residential and small-commercial air-conditioners and heat pumps.


Due to the diversity and complexity of protection schemes in induction motor loads \cite{TDC}, developing adequate aggregate protection models and setting proper parameters is a complex and challenging task. The fractions of the motor loads in the composite load model vary based on different regions, seasons and day types. In addition, for different building types, the motor types vary significantly with corresponding protection schemes \cite {DanReport}. Therefore, a composite protection model is needed to aggregate the performance of the protection of all the motor loads in the composite load model. Recent works have looked into the composite protection behavior of residential and commercial building motor loads \cite{TDC,ISGT18,PESGM18,ISGT19}. In \cite{ISGT18}, the authors presented a methodology to generate composite protection profiles for different commercial buildings in representative cities across different climate zones. Using integrated transmission and distribution dynamic co-simulation with detailed models of motor loads and associated protection schemes, the authors showed in \cite{TDC} that the aggregate protection response of
motor loads of different types can vary significantly, which were not adequately captured or reflected by the existing protection in WECC composite load model.

While the composite protection behavior of motor loads can be complex and time-varying, there is a need to simplify this model for easier integration into higher-level (transmission systems) simulation programs. Moreover, even though the distribution networks are seeing an increased deployment of advanced sensors and meters, it is reasonably expected that motor load fractions will likely be unknown, with their estimates being available with associated estimation errors. This paper uses a recently introduced mathematical abstraction (in \cite{ACM19}) to represent the protection profiles, and proposes a nonlinear regression problem to obtain a suitable simplified model of the composite protection profile. Section\,\ref{S:methods} of this paper introduces a mathematical abstraction of the protection schemes. Section\,\ref{S:problem} presents the nonlinear regression problem to obtain the parameters of a suitably simplified composite protection scheme. Illustrative results on the application of the framework are presented in Section\,\ref{S:results}, along with a study of the accuracy of the protection model under various uncertainty scenarios. The article is concluded in Section\,\ref{S:concl}. 

\section{Protections: Mathematical Modeling}\label{S:methods}

\begin{figure}[thpb]
      \centering
	\includegraphics[scale=0.2]{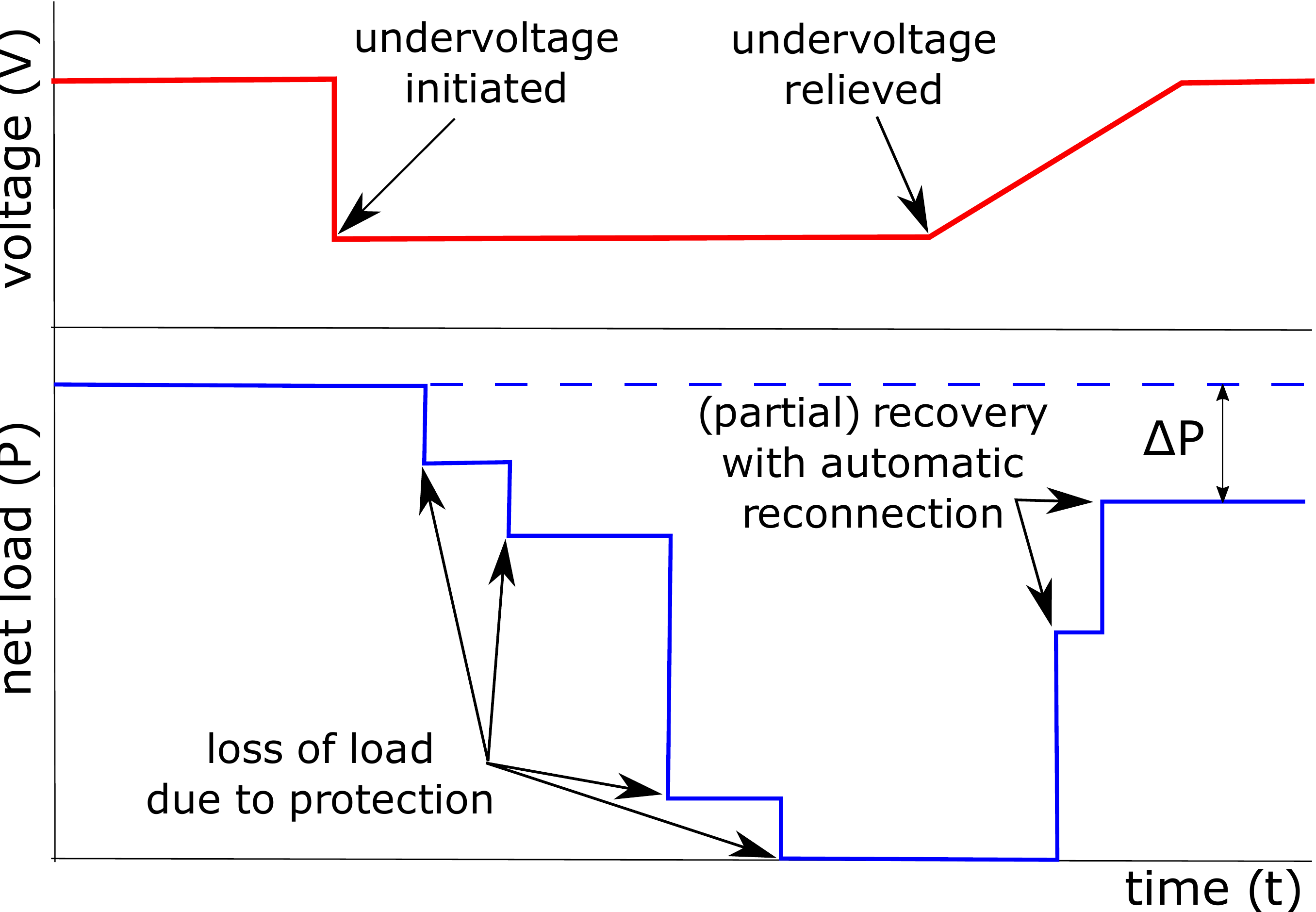}
      \caption{Typical load tripping profile.}
      \label{F:tripping}
   \end{figure}

Motors are typically protected by multiple devices, such as relays, contactors, thermal protection, etc. During a fault, as the voltage drops below a certain limit for longer than a certain duration, multiple protection mechanisms could be triggered to trip the associated motor load. Fig.\,\ref{F:tripping} illustrates how an aggregate motor load may respond during a voltage event due to the various protection schemes activated over the duration of the fault (Note that Fig.\,\ref{F:tripping} ignores the motor dynamics, but focuses only on the effect of the protection). 
Understanding the behavior of motor loads under the action of different protection schemes is of paramount importance. 

Modeling protection schemes, in general, is a challenging task. The protection equipment present in different motors vary widely in their operating parameters (i.e. tripping and reconnection behavior). Furthermore, the response parameters of a protection device may not be static, and can also depend on factors such as the loading on the motor (e.g. fully loaded motors will likely trip earlier than lightly loaded motors), which may in turn depend on conditions such as the outside air temperature, occupancy of a buildings, etc. In this paper, we adopt the mathematical model of the protection schemes introduced recently \cite{ACM19}, which determines, given a certain fault, if the protection would be tripped or if it would remain in the operational region based on some static trip conditions. 

\subsection{Modeling Protection schemes}\label{S:individual}

\begin{definition}
 \cite{ACM19} Trip-zone for a given protection scheme-$i$, denoted by $\mathcal{T}^i$\,, is defined as the set of pairs of voltage levels at fault ($v_f$) and the fault duration values ($t_f$) such that the protection-$i$ is \textit{tripped} if and only if $(\tau_f,v_f)\in\mathcal{T}^i$\,, i.e.
\begin{align*}
(\tau_f,v_f)\in\mathcal{T}^i \iff \text{protection-$i$ is \textit{tripped}.}
\end{align*}
\end{definition}

  Each protection scheme can be modeled mathematically in the form of a discrete-valued function $f^i:\mathbb{R}_{\geq0}^2\mapsto\lbrace 0,1\rbrace$ defined as follows:
  \begin{align}
  f^i(\tau_f,v_f) = \left\lbrace\begin{array}{ll}0,& (\tau_f,v_f)\in\mathcal{T}^i\\
1, & \text{otherwise}\end{array}\right.
  \end{align}
 where the value of the function is 0 whenever the protection is tripped (i.e. the motor is disconnected from the network), and 1 when the protection has not been tripped (i.e. the motor is still connected to the network). Note that the shape of the trip-zone is different for different protections.
 
 \begin{remark}
 In this work, we focus only on the tripping of the protection and not on the reconnection event. 
 \end{remark}
 
 %
%
%

\begin{figure}[thpb]
      \centering
	\includegraphics[scale=0.4]{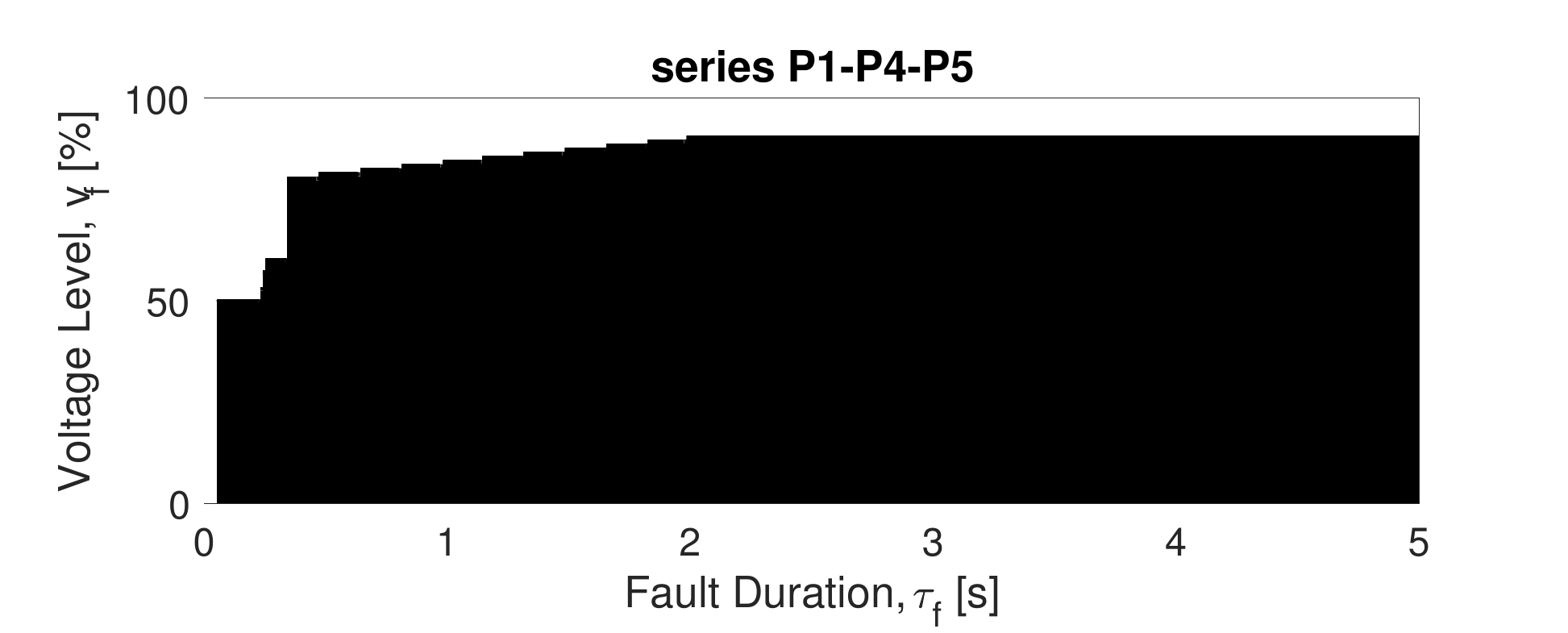}
      \caption{Examples of a protection diagram for the protection scheme P1-P4-P5. Black region denotes the trip-zone..}
      \label{F:P1P4P5}
   \end{figure}


   
Motor protection schemes commonly found in commercial buildings in United States can be categorized into five different types, each of which is characterized by a range of voltage deviations and durations for tripping after the fault - 1) electronic relays (P1), 2) current overload relays (P2), 3) thermal protection (P3), 4) contactors (P4) and 5) building management system (P5). For more details readers are referred to \cite{DanReport,ISGT18}. Most often the motors are protected not by a single mechanism, but by a combination of the different protection schemes, in a series combination. In a series combination of protections, each protection needs to be in operational state in order for the motor to be connected to the network. Let us consider some protection-$k$ which is a series combination of protection-$i$ and protection-$j$\,. Then the following holds:
\begin{subequations}
\begin{align}
\!\!\mathcal{T}^k&=\mathcal{T}^i\cup\mathcal{T}^j,\\
\!\!\text{(or, equivalently)}~~ f^k(\tau_f,v_f)&=f^i(\tau_f,v_f)\cdot f^j(\tau_f,v_f),
\end{align}\end{subequations}
i.e. the trip-zone of a series combination is a union of the trip-zones of each of the protections in the combination. In other words, the motor is disconnected from the network whenever any of the protections in the series combination trips. Fig.\,\ref{F:P1P4P5}, adopted from \cite{ACM19}, shows an example of the protection diagram for the series combination P1-P4-P5.

\subsection{Generating Composite Protection}\label{S:composite}

Different induction motors are protected by various (series combinations of) protection schemes. Let us denote the set of all available (series) combinations of protection schemes by 
\begin{align*}
\mathcal{P}:\,\text{set of all available protection combinations}\,.
\end{align*}
such that each member of the set $\mathcal{P}$ is unique. Composite protection modeling is about constructing a reduced order protection model that can predict the fractions of total motor-load tripped during a fault.
\begin{definition}
\cite{ACM19} The composite protection scheme of a collection of motors served by (combinations of) protection schemes belonging to the set $\mathcal{P}$ can be mathematically modeled in the form of the (discrete-valued) function $F:\mathbb{R}_{\geq 0}^2\mapsto[0,1]$ defined as follows:
\begin{align}\label{E:composite_true}
F(\tau_f,v_f)=\sum_{i\in\mathcal{P}}\pi^i\,f^i(\tau_f,v_f)
\end{align}
where $\pi^i\in[0,1]$ is the fraction of the motor-load served by protection-$i$\,, i.e. $\sum_{i\in\mathcal{P}}\pi^i=1$. Henceforth $F$ is referred to as the `composite protection function'.
\end{definition}
\begin{remark}
Note that the fractions of the motor-load served by a particular protection type is a time-varying quantity. Thus the composite protection function will also be time-varying. For the purpose of this work, we do not explicitly model the time variability, while noting that the approach extends to the time-variable composite protection functions as well.
\end{remark}
 
$F$ takes discrete values between $0$ and $1$\,, with the value of $0$ referring to all the motor-loads are disconnected, while the value of $1$ refers to all motor-loads being connected. In recent work \cite{ISGT18}, the authors presented a methodology to approximate the motor-load fractions ($\pi^i$) for each protection combination based on typical commercial buildings' (hourly) energy consumption profiles, in different climate-zones. Composite protection schemes can be used to predict the fractions of motor-loads that will be tripped during a fault (interested readers are referred to Example\,1 in \cite{ACM19}).


\section{Composite Protection Modeling}\label{S:problem}

\begin{figure}[thpb]
      \centering
	\includegraphics[scale=0.2]{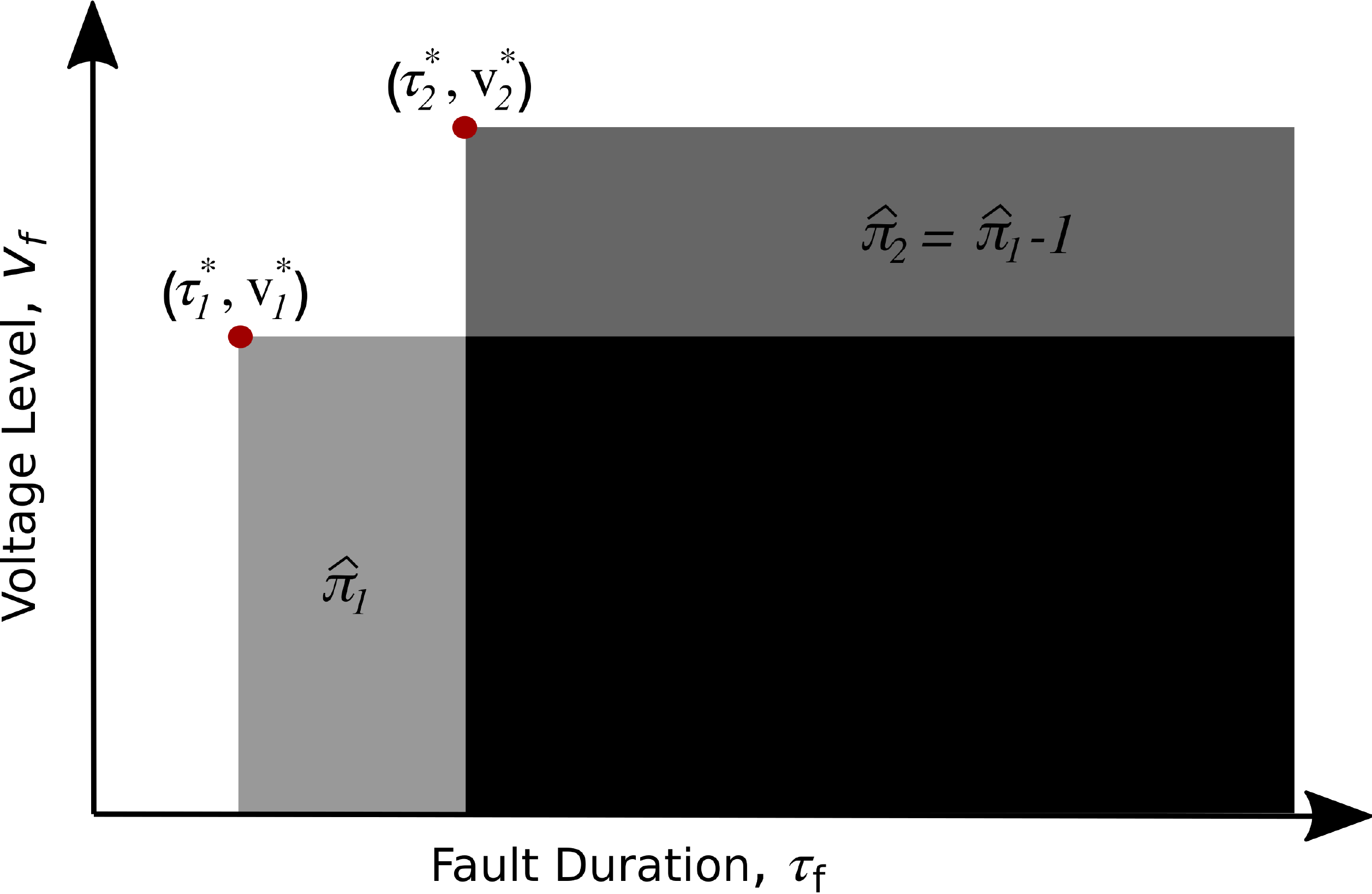}
      \caption{Simplified model of the composite protection scheme.}
      \label{F:target}
   \end{figure}   

  The composite protection scheme can be quite complex with rather arbitrarily shaped trip-zones. While such detailed models can be quite useful for understanding the behavior in the distribution networks, these are not very easy to integrate with composite transmission-distribution studies. A simplified model with reduced complexity appears to be necessary, which approximates the detailed composite protection model \textit{as best as possible}. Readers are referred to WECC composite load modeling efforts (\cite{composite}, and related works) for more details. In this paper, our goal is to approximate the composite protection model using the simplified form as follows (Fig.\,\ref{F:target}):
  \begin{subequations}\label{E:composite_approx}
  \begin{align}
  \!\!\!\!\widehat{F}(\tau_f,v_f)&\!:=\hat{\pi}_1\,\widehat{F}_1(\tau_f,v_f)+\hat{\pi}_2\,\widehat{F}_2(\tau_f,v_f)\!\\
\!\!\!\!\forall i\!\in\!\lbrace 1,2\rbrace\!\!:~\widehat{F} _i(\tau_f,v_f)&\!:=\! \left\lbrace\! \begin{array}{cl}0, & \tau_f\geq\tau_i^*~\&~v_f\leq v_i^*\\
1, & \text{otherwise}\end{array}\!\right.\!\\
1&\!=\hat{\pi}_1+\hat{\pi}_2\,.
  \end{align}\end{subequations}
  Here $\widehat{F}_1$ ($\widehat{F}_2$) denotes the protection scheme that serves $\hat{\pi}_1$ ($\hat{\pi}_2$) fraction of motor-loads, with a trip-zone that is parameterized by a trip-voltage $v^*_1$ ($v^*_2$) and trip-duration $\tau^*_1$ ($\tau^*_2$). The goal is to find the parameters 
  \begin{align}
  \Phi=\lbrace \hat{\pi}_1,\tau_1^*,v_1^*,\hat{\pi}_2,\tau_2^*,v_2^*\rbrace
  \end{align}
  such that the simplified protection scheme $\widehat{F}$ in \eqref{E:composite_approx} approximates the true protection scheme $F$ in \eqref{E:composite_true}. We set up a nonlinear regression problem to find the parameters $\Phi$ that gives the \textit{best} approximation ($\widehat{F}$) of the \textit{true} composite protection ($F$). This is done in the following steps:
  \begin{enumerate} 
  \item Randomly select $N$ points from the $(\tau_f,v_f)$-space and note down the values of the \textit{true} composite protection function $F$ at those points (from \eqref{E:composite_true}). Let us denote these points by $(\tau_f^j,v_f^j)$ and the corresponding value of $F$ as $y^j=F(\tau_f^j,v_f^j)$\,, for each $j\in\lbrace 1,2,\dots,N\rbrace$\,.
  \item Construct the \textit{cost function} as 
  \begin{align}
  J(\Phi):=\frac{1}{2N}\sum_{j=1}^N\left(\widehat{F}(\tau_f^j,v_f^j)-y^j\right)^2
  \end{align}
  \item Solve the following optimization problem:
  \begin{subequations}\label{E:optimization}
  \begin{align}
  \min_{\Phi}~& J(\Phi)\\
  \text{s.t.}\quad & v_i^*\in[0,100]\,,\,\tau_i^*\in[0,5]~\forall i\in\lbrace 1,2\rbrace,\,\\
  					& \hat{\pi}_1+\hat{\pi}_2=1\,.
  \end{align}\end{subequations}
  \end{enumerate}
  Note that the optimization problem \eqref{E:optimization} cannot be solved directly in the present form, since it involves functions ($\widehat{F}_{1,2}$) that are described in conditional forms \eqref{E:composite_approx}. We overcome this problem by using \textit{logistic functions} to model the protection functions $\widehat{F}_{1,2}$\,. \textit{Logistic functions} $h:\mathbb{R}\mapsto[0,1]$ are approximations of \text{step functions} and are defined as follows:
  \begin{align}
  \text{(logistic)}\quad h(x;\alpha):=\frac{1}{1+\exp(-\alpha\,x)}
  \end{align}
  where $\alpha>0$ is a \textit{steepness} parameter related to the slope of the function at $x=0$\,. The functions $\widehat{F}_{1,2}$ are approximated using logistic functions as follows:
  \begin{align}\label{E:F_logistic}
  \widehat{F}_i(\tau_f,v_f)=1-h(\tau_f-\tau_i^*;\alpha_{\tau})\left(1-h(v_f-v_i^*;\alpha_v)\right)
  \end{align}
  for each $i\in\lbrace 1,2\rbrace$\,, for some chosen $\alpha_\tau,\alpha_v>0$\,. The optimization problem \eqref{E:optimization} is solved via IPOPT \cite{Wachter:2006} using the logistic functional representation of $\widehat{F}_{1,2}$ in \eqref{E:F_logistic}.
  
  Ideally, one would like to solve \eqref{E:optimization} with as many data points as possible (large $N$), however, due to computational limitations $N$ has to be reasonably small. Thus the data points need to be selected judiciously so that the approximation is sufficiently accurate. Typically, the protection function changes value rapidly when $\tau_f$ is near $0$\,s. Moreover, during faults $v_f$ is typically close to $50$\,\%. Therefore we select the data points by assigning some weights $w(\cdot)\in[0,1]$ to every point on the $(\tau_f,v_f)$-axis as follows:
  \begin{align*}
  w(\tau_f,v_f)=1-\left(1-e^{-\beta_\tau\tau_f}\right)\left(1-e^{-\beta_v\left(v_f-50\right)}\right),
  \end{align*}
  for some $\beta_\tau,\beta_v>0$\,, and selecting $N$ points randomly from all points that have larger than a chosen weight. 
  
 MAE)  Finally, we measure the accuracy of the approximation using the following mean absolute error (metric:
  \begin{align}
  \text{(MAE)}\quad \epsilon:=\frac{1}{M}\sum_{k=1}^M\left|\widehat{F}(\tau_f^k,v_f^k)-F(\tau_f^k,v_f^k)\right|
  \end{align}
  where the $M$($\gg\!N$) points $(\tau_f^k,v_f^k)$ are selected randomly (and separately from the data points used in \eqref{E:optimization}) using, say, \textit{Latin hypercube sampling} technique from the $(\tau_f,v_f)$-space.
  
\section{Numerical Results}\label{S:results}
  %
\begin{figure}[thpb]
      \centering
	\includegraphics[scale=0.4]{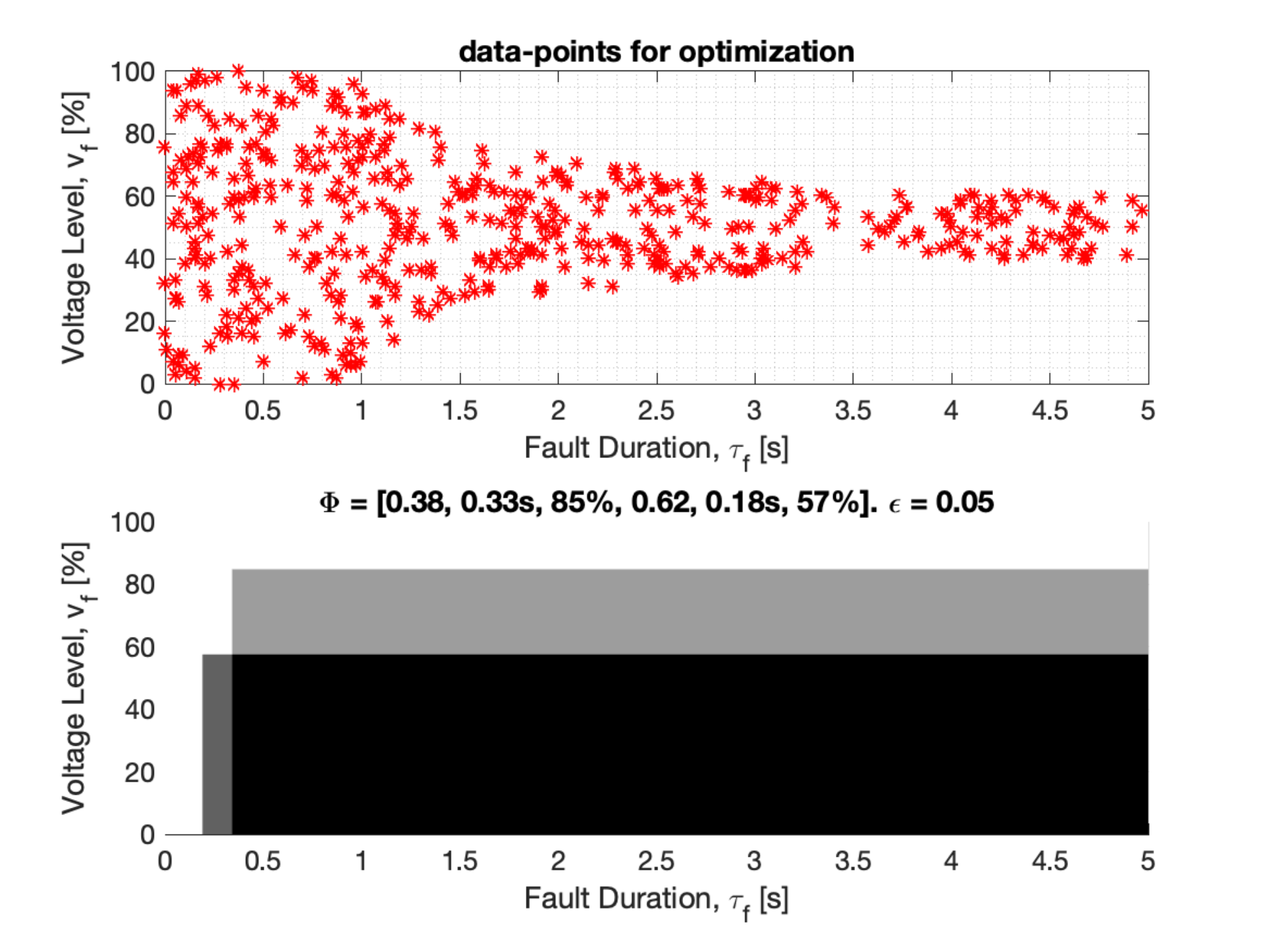}
      \caption{Simplified composite protection for Example\,1 in \cite{ACM19}.}
      \label{F:example_error}
   \end{figure}   
\begin{figure*}[thpb]
\centering
\captionsetup{justification=centering}
\hspace{-0.37in}\subfigure[Motor A protection]{
\includegraphics[scale=0.33]{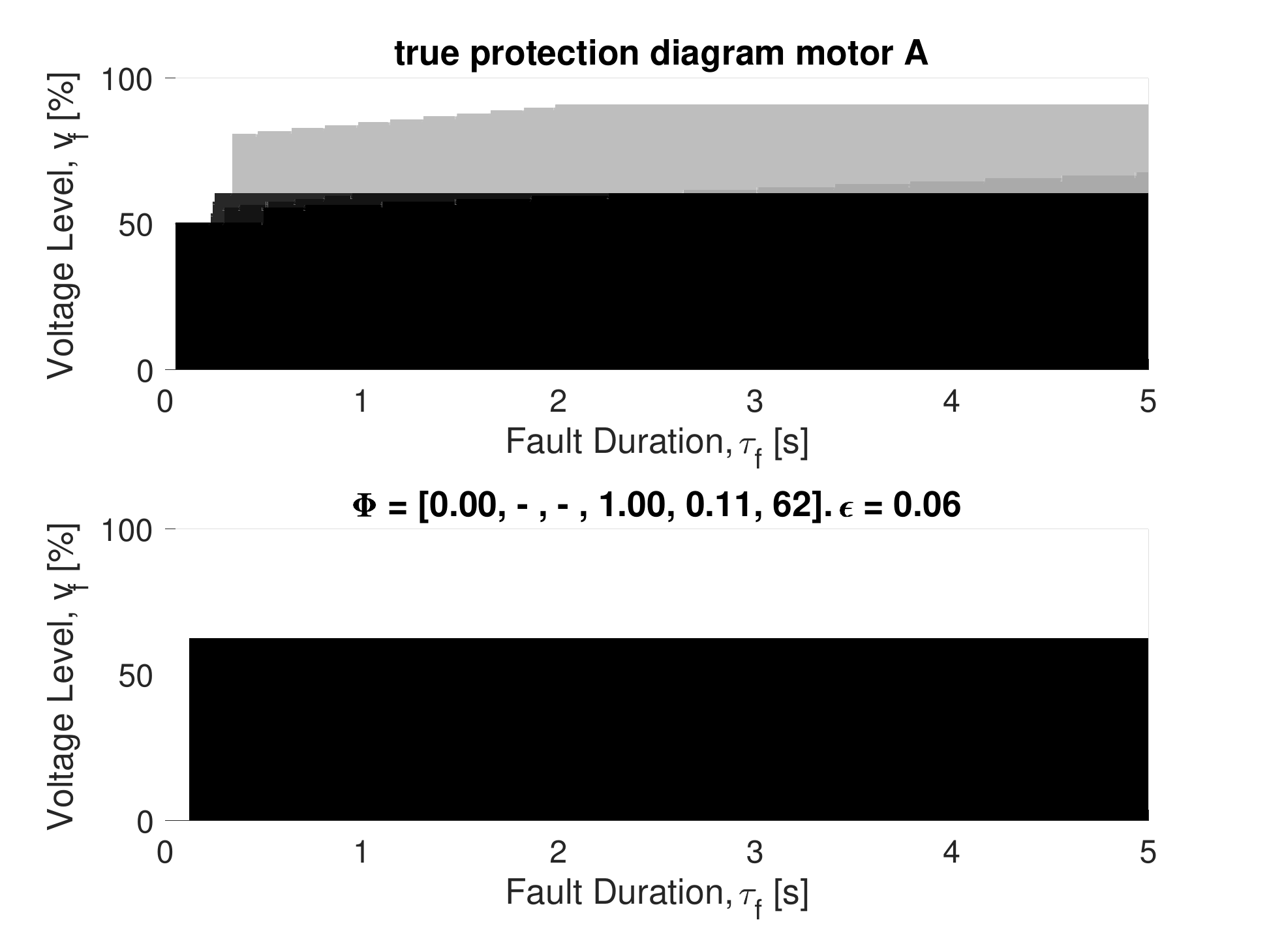}\label{F:test_A}
}
\hspace{-0.37in}
\subfigure[Motor C protection]{
\includegraphics[scale=0.33]{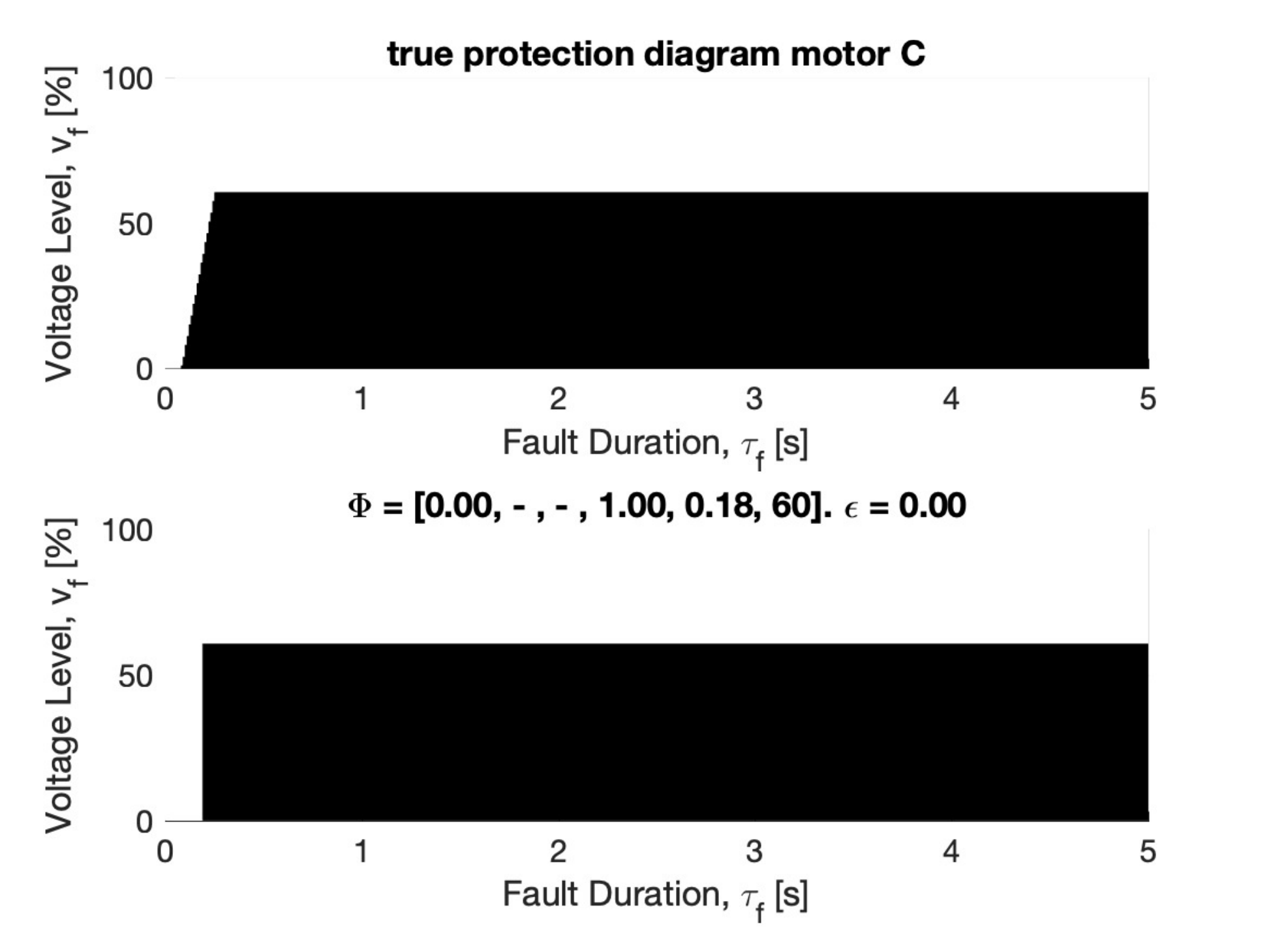}\label{F:test_C}
}
\hspace{-0.37in}
\subfigure[Motor D protection]{
\includegraphics[scale=0.33]{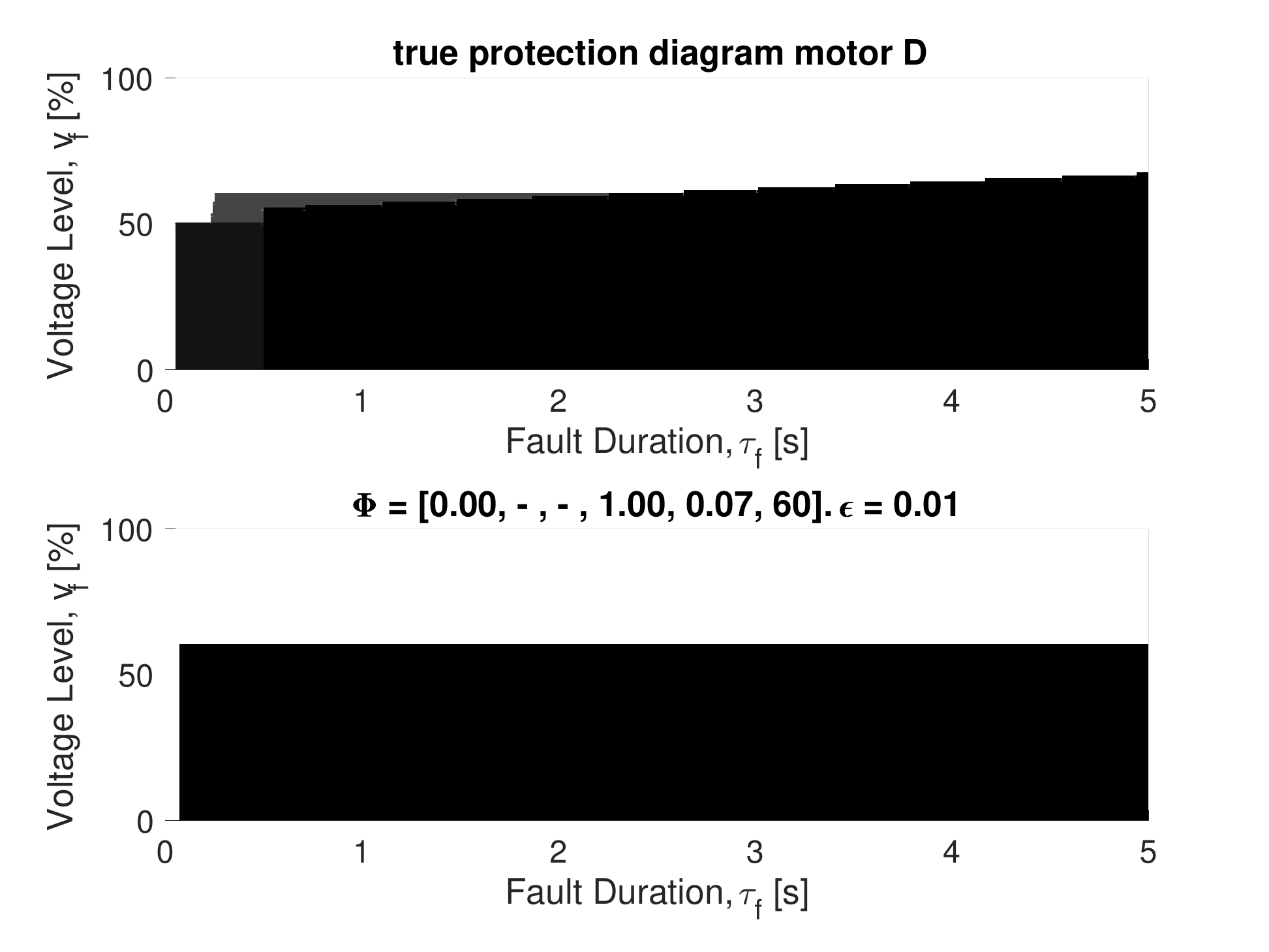}\label{F:test_D}
}\hspace{-0.37in}
\caption[Optional caption for list of figures]{True and simplified composite protection schemes for the test-case in \cite{PESGM18}. Load fractions are listed in Table\,\ref{Tab:params}.}
\label{F:test}
\end{figure*}
\begin{figure}[thpb]
      \centering
	\includegraphics[scale=0.4]{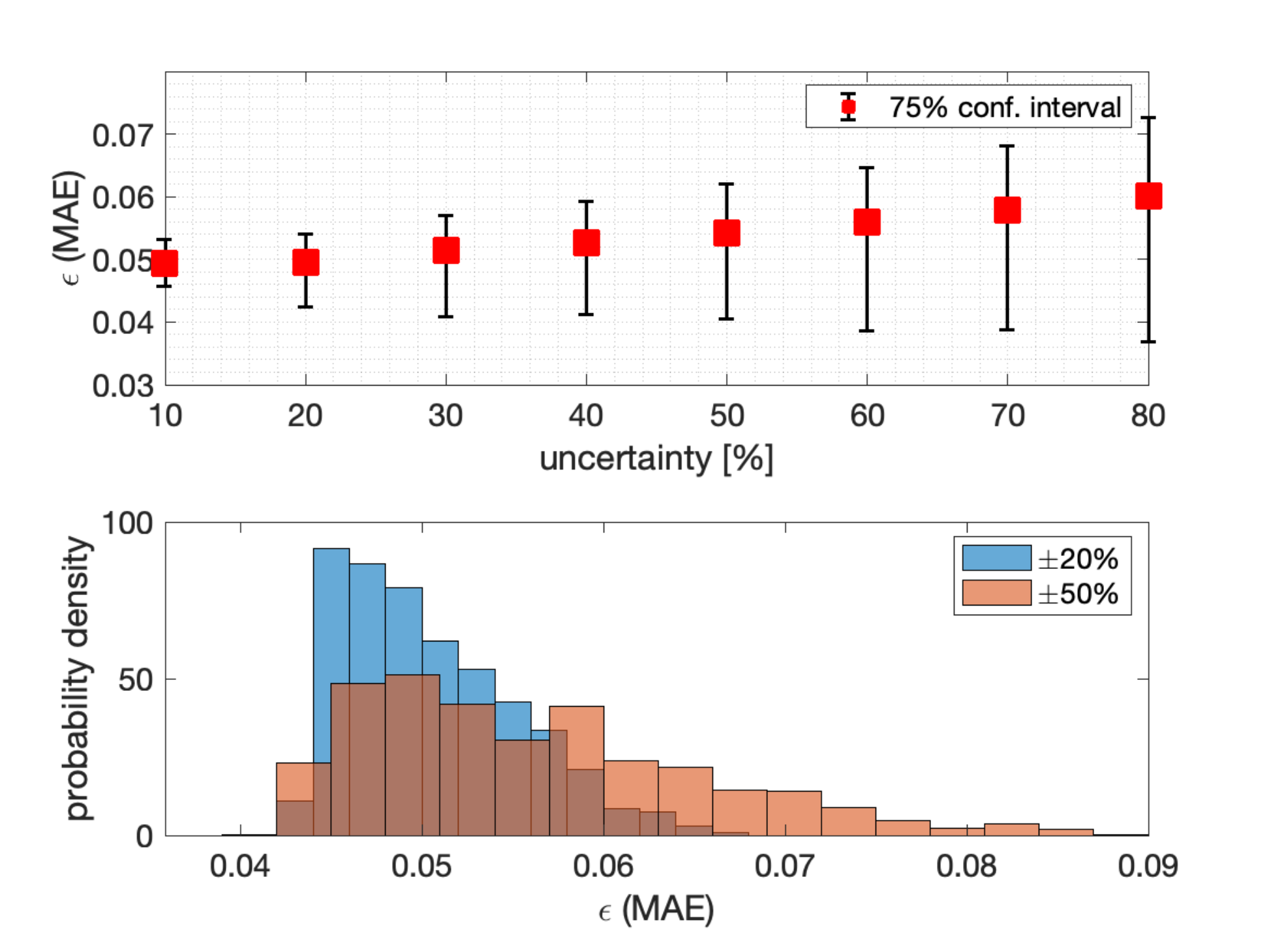}
      \caption{Mean absolute error (MAE) statistics under uncertainties in motor-load fractions for Example\,1 of \cite{ACM19}.}
      \label{F:error_all}
   \end{figure}
\begin{figure}[thpb]
      \centering
	\includegraphics[scale=0.4]{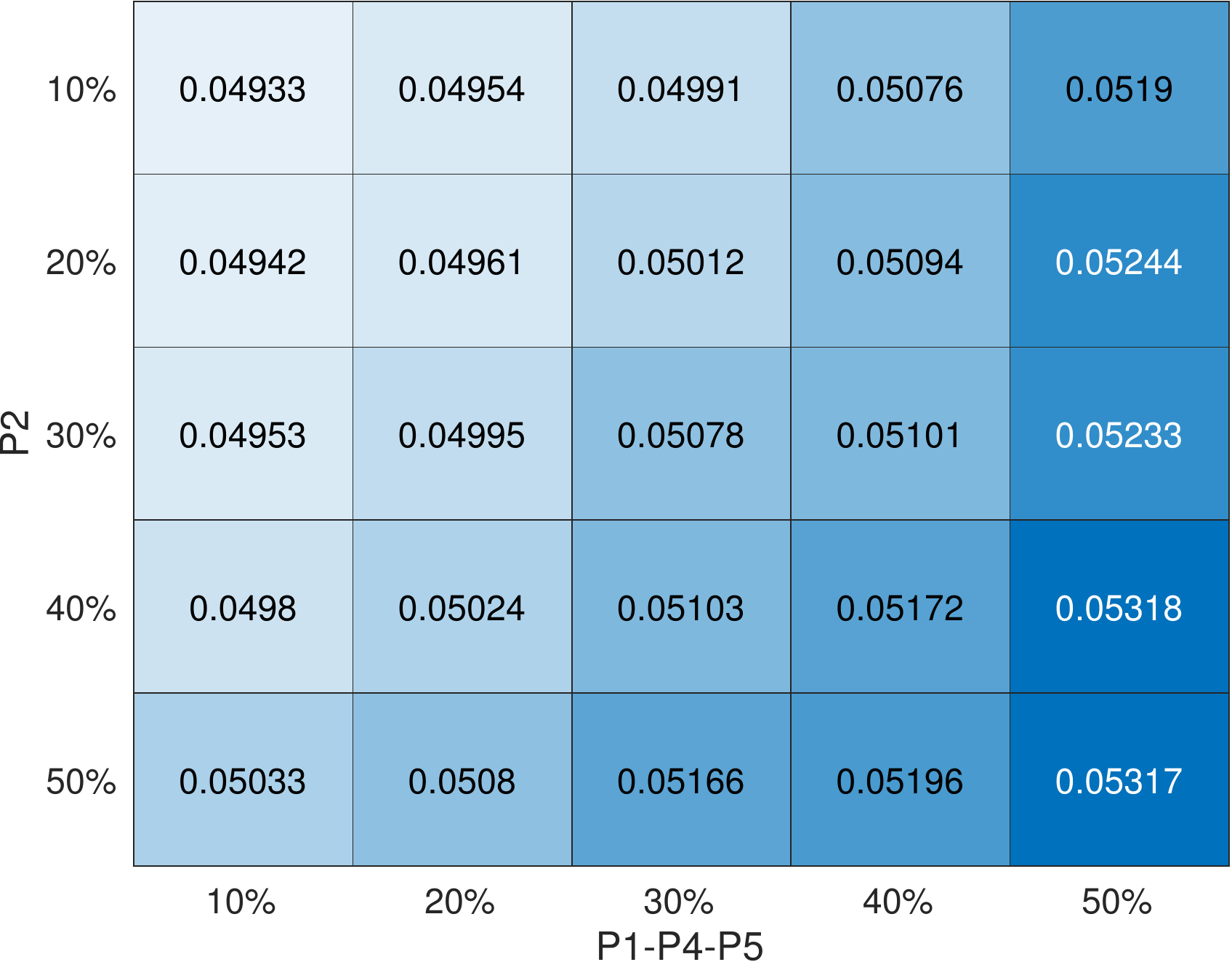}
      \caption{Average MAE values under uncertainties in P2 and P1-P4-P5 fractions (Example\,1 in \cite{ACM19}).}
      \label{F:error_mat}
   \end{figure}

   We apply the aforementioned optimization framework to the problem in Example\,1 of \cite{ACM19}, to obtain the simplified composite protection diagram. The result is shown in Fig.\,\ref{F:example_error}, where the top plot shows the selected data-points for \eqref{E:optimization} and the bottom plot shows the resulting simplified protection scheme with a mean absolute approximation error ($\epsilon$) of 0.05. However, in practice, the information needed for the optimization problem are often unknown. As such the accuracy of the simplified model is dependent on the accuracy of those information. We evaluate the performance of the simplified model under uncertainties in the motor-load fractions. Fig.\,\ref{F:error_all} shows the MAE statistics when we allow all the motor-load fractions corresponding to protection schemes P1, P2, P3, P5 and P1-P4-P5 to have varied level of uncertainties modeled as below:
   \begin{align*}
   \pi^\text{actual}=(1+\gamma)\,\pi^\text{optimization}\,,
   \end{align*}
   i.e. the fraction used in optimization is different from the actual one. Value of $\gamma$ is varied between $\pm10\%$ to $\pm80\%$\,. The mean MAE value and its 75\% confidence interval increases as the uncertainty goes higher. The probability distributions of MAE for uncertainty levels $\pm20\%$ and $\pm50\%$ are shown to further illustrate this observation. Fig.\,\ref{F:error_mat} shows the mean MAE values when the uncertainties are introduced to only the motor-load fractions corresponding to protections P2 and P1-P4-P5 (while other motor-loads fractions being unperturbed).
   
   Next we consider the test-cases developed in \cite{PESGM18}. In particular, we apply the optimization framework to obtain simplified composite protection functions for motor loads in a hotel, large retail, medium retail, school, warehouse and supermarket. The protection schemes and the associated motor-load fractions used in the study are listed in Table\,\ref{Tab:params}\,. Note that, out of a total of 31 possible combinations of protection schemes, only seven were found in the buildings considered (based on the study done in \cite{DanReport,ISGT18}). The optimization problem was run separately for the motor types A, B, C and D to obtain their simplified composite protection schemes. The results are shown in Fig.\,\ref{F:test} (due to similarity between the protection schemes of motor A and B, only A is shown in the plot). In this particular case, it turned out that the $\hat{\pi}_1=0$ for all the motors (but not expected in general, e.g. Fig.\,\ref{F:example_error}).
   

   \begin{table}[thpb]
\caption{Test-case Protection Parameters}
\label{Tab:params}
\centering
\begin{tabular}{|*{5}{c|}}\hline
$\mathcal{P}$ & $\pi$ (for motors A,\,B,\,C\,\&\,D)\\\hline
P3 & $\lbrace 0.00,0.00,0.00,0.08\rbrace$ \\\hline
P2-P4 & $\lbrace 0.09,0.08,0.00,0.00\rbrace$\\\hline
P3-P4 & $\lbrace 0.08,0.00,0.00,0.20\rbrace$\\\hline
P2-P5 & $\lbrace 0.00,0.00,1.00,0.00\rbrace$\\\hline
P1-P4-P5 & $\lbrace 0.25,0.21,0.00,0.00\rbrace$\\\hline
P2-P4-P5 & $\lbrace 0.58,0.69,0.00,0.00\rbrace$\\\hline
P3-P4-P5 & $\lbrace 0.00,0.02,0.00,0.72\rbrace$\\\hline
\end{tabular}
\end{table}

\section{Conclusions and Future Work}\label{S:concl}
There is a need for high-fidelity composite load protection models for induction motor loads to better represent the aggregate dynamic behavior of distribution systems in the transmission system dynamic simulations and studies. This work builds on recent developments on the composite protection modeling, to propose an optimization framework to generate simplified composite protection schemes. Introducing a mathematical abstraction of the protection schemes, a nonlinear regression problem is formulated to suitably approximate complex protection schemes in a simple parametric form. Numerical results are presented to illustrate the application of the framework. Future studies will focus on validation of the simplified models through detailed transmission-distribution co-simulations, as well as on the sensitivity of the solution to uncertainties in load compositions.




%

\end{document}